\documentclass[onecolumn]{mn2e}
\newcommand{\be}{\begin{eqnarray}}
\newcommand{\ee}{\end{eqnarray}}

\newcommand{\Ms}{M_{\star}}
\newcommand{\Msun}{{\rm M}_{\odot}}
\newcommand{\Mp}{M_p}

\newcommand{\ME}{{\rm M}_{\oplus}}
\newcommand{\RE}{{\rm R}_{\oplus}}
\newcommand{\mey}{{\rm M}_{\oplus} \; {\rm yr}^{-1}}

\usepackage{graphicx}
\usepackage{amsmath}
\graphicspath{{figures/}}

\begin{document}

\title[On the formation of the Kepler--10
planetary system]{On the formation of the
  Kepler--10 planetary system}

\author[Caroline Terquem] { Caroline Terquem \\ Physics Department,
  University of Oxford, Keble Road, Oxford OX1 3RH, UK \\
  Institut d'Astrophysique de Paris, UPMC Univ Paris 06, CNRS,
  UMR7095, 98 bis bd Arago, F-75014, Paris, France \\ E-mail:
  caroline.terquem@astro.ox.ac.uk }

\date{}

\pagerange{\pageref{firstpage}--\pageref{lastpage}} \pubyear{}

\maketitle

\label{firstpage}

%
%

\begin{abstract}
  In this paper, we investigate the conditions required for the 3 and
  17~$\ME$ solid planets in the Kepler--10 system to have formed
  through collisions and mergers within an initial population of
  embryos.  By performing a large number of $N$--body simulations, we
  show that the total mass of the initial population had to be
  significantly larger than the masses of the two planets, and that
  the two planets must have built--up farther away than their present
  location, at a distance of at least a few~au from the central star.
  The planets had to grow fast enough so that they would detach
  themselves from the population of remaining, less massive, cores and
  migrate in to their present location.  By the time the other cores
  migrated in, the disc's inner edge would have moved out so that
  these cores cannot be detected today.  We also compute the critical
  core mass beyond which a massive gaseous envelope would be accreted
  and show that it is larger than 17~$\ME$ if the planetesimal
  accretion rate onto the core is larger than $10^{-6}$~$\mey$.  For a
  planetesimal accretion rate between $10^{-6}$ and $10^{-5}$~$\mey$,
  the 17~$\ME$ core would not be expected to have accreted more than
  about 1~$\ME$ of gas.  The results presented in this paper suggest
  that a planetary system like Kepler--10 may not be unusual, although
  it has probably formed in a rather massive disc.
\end{abstract}


%
%

\section{Introduction}
\label{sec:intro}

Since the detection of the first rocky extrasolar planet (Corot~7b,
Queloz et al. 2009, L\'eger et al. 2009), a large number of similar
objects have been observed by {\em Kepler} (Borucki et al. 2011,
Batalha et al. 2013).  As most of the planets detected by {\em Kepler}
have not been confirmed by radial velocity measurements, the mass is
not in general available and we have to rely on models linking the
radius to the mass to classify the planets.  Bucchave et al. (2014)
and Marcy et al. (2014) have proposed that objects with radii smaller
than $\sim$ 1.5 Earth radius ( $\RE$), between $\sim$ 1.5 and $\sim$ 4
$\RE$ and larger than $\sim$ 4 $\RE$ are, respectively, terrestrial
planets, planets with a rocky core and a hydrogen--helium envelope,
and ice or gas giants.  According to this classification, the planet
Kepler--10c, with a radius of 2.35~$\RE$, is expected to have a
gaseous envelope.  Yet, its mass has been determined by radial
velocity measurements, and being about 17~$\ME$, it indicates that the
planet has a very high density of 7~g~cm$^{-3}$ and is likely to be
solid (Dumusque et al. 2014).

Solid mass planets are believed to be formed through a process
starting with the sedimentation and collisional growth of dust grains
in a protostellar disc, followed by solid body accretion of km--sized
objects (Lissauer 1993, Papaloizou \& Terquem 2006 and references
therein) or cm--sized pebbles (Lambrechts \& Johansen 2012).  The
formation of massive solid cores, which are the nucleus of gas giant
planets, is believed to occur through collisions (also called giant
impacts) between embryos.

Once the planets reach a mass on the order of a tenth of an Earth
mass, they start migrating in the disc on a timescale comparable to or
smaller than the planet formation timescale (Ward 1997).  Recent
hydrodynamical simulations (Pierens, Cossou \& Raymond 2013) have
shown the difficulty of forming very massive cores through giant
impacts of terrestrial mass planets.  Because of migration, the
evolution of a population of such planets tend indeed to result in a
resonant chain rather than in a single massive core (see also Terquem
\& Papaloizou 2007).  Very massive cores are found only when starting
with a population of planets of at least 2--3~$\ME$.  Alternatively,
massive cores could form by continuous accretion of planetesimals, but
the timescale for forming a $\sim$~10~$\ME$ core is usually found to
be longer than the migration timescale (see Tanigawa 2008 and
references therein).

The planetary system Kepler--10, which comprises at least two planets,
harbours the first rocky planet that was discovered by {\em Kepler}.
Radial velocity measurement from Keck--HIRES, made immediately after
the detection by {\em Kepler}, enabled the mass of Kepler--10b to be
determined (Batalha et al. 2011).  More recent observations from
HARPS--N have improved the precision on the mass of Kepler--10b, and
have allowed the determination of the mass of Kepler--10c: the system
has a super Earth of 3.3~$\ME$ at 0.017~au, and a Neptune--mass planet
of 17.2~$\ME$ at 0.24~au (Dumusque et al. 2014).  With a radius of
2.35~$\RE$, the Neptune--mass planet therefore has a very high
density.  It is the first known solid planet with a mass above
10~$\ME$ (Kepler--131b may be similar to Kepler--10c, but its mass has
not yet been determined with certainty, Marcy et al. 2014).  The fact
that Kepler--10c is solid has come as a surprise, as it is commonly
believed that the critical core mass, above which accretion of a
massive gaseous envelope occurs, is $\sim 10$~$\ME$.

In this paper, we investigate the conditions required for two planets
similar to those in the Kepler--10 system to form through collisions
and mergers within an initial population of embryos
(section~\ref{sec:formation}).  We show that the total mass of the
initial population has to be significantly larger than the masses of
the two planets, and that the two planets must have built--up farther
away than their present location, at a distance of at least a few~au
from the star.  We then compute the critical core mass at the location
where the Neptune--mass planet formed
(section~\ref{sec:criticalmass}).  We find that it is larger than
17~$\ME$ if the planetesimal accretion rate onto the core is larger
than $10^{-6}$~$\mey$.  We finally discuss our results in
section~\ref{sec:discussion}.

%
%

\section{Formation of massive solid planets}
\label{sec:formation}


In this section, we investigate scenarii that could result in a
planetary system like Kepler--10, comprising two solid planets of
about 3 and 17~$\ME$ at 0.017 and 0.24~au, respectively.

\subsection{In--situ formation}

\label{sec:insitu}

Let us first consider whether the planets could have formed {\em in
  situ}.  An embryo at 0.017 or 0.24~au from the star could in
principle grow through accretion of solid material in the form of
either dust, planetesimals or solid cores.  However, {\em in situ}
growth can only happen if the embryo is prevented from migrating onto
the central star, i.e. if its orbit is inside the disc's inner edge.
Loss of contact with the disc then makes it difficult for the embryo
to accrete dust or planetesimals migrating within the disc towards the
star.  The orbit of more massive cores also migrating in could in
principle cross that of the embryo, resulting in collisions and
growth.  However, as we will see in this section, incoming cores tend
to be captured in mean motion resonances rather than collide with
cores already within the disc's inner egde.  It is therefore unlikely
that the planets in the Kepler--10 system have formed {\em in situ}.

We have assumed in the above discussion that the embryo would
  stop migrating after entering the cavity.  However, Masset et al.
  (2006) have suggested that cores would be trapped at the edge of the
  disc, rather than penetrating inside the cavity, due to the effect
  of the corotation torque.  In this context, the embryo would not
  lose contact with the disc and could continue to accrete dust and/
  or planetesimals migrating within the disc.  However, it is not
  clear that trapping of the cores would happen in the presence of MHD
  turbulence, which is likely to have been present in the disc at the
  location of the planets in the Kepler--10 system.  Whether the disc
  keeps the planet trapped or not depends strongly on the profile of
  the surface density at the edge (Masset et al. 2006).  Also, recent
  MHD simulations indicate that planets with masses as small as $\sim
  10$~$\ME$ can open up gaps in turbulent regions of discs with net
  vertical magnetic flux (Zhu, Stone \& Rafikov 2013).  The corotation
  torque acting on such planets would be much reduced, so that
  trapping would not occur.

We now investigate whether the dynamical evolution of a population of
cores migrating inwards within the disc can result in the formation of
a super Earth (with a mass of a few Earth masses) and a massive solid
planet (with a mass similar to that of Neptune) at 0.017 and 0.24~au,
respectively.

\subsection{Numerical integration}

To compute the evolution of a population of cores migrating through a
disc, we use the $N$--body code described in Papaloizou~\& Terquem~(2001)
in which we have added the effect of the disc torques (see also
Terquem \& Papaloizou 2007).

The equations of motion for each core are:

\begin{equation} {d^2 {\boldsymbol{r}}_i\over dt^2} = -{GM_\star
    \boldsymbol{r}_i \over |\boldsymbol{r}_i|^3} -\sum_{j=1\ne i}^N
  {G M_j \left(\boldsymbol{r}_i- \boldsymbol{r}_j \right) \over 
     |\boldsymbol{r}_i- \boldsymbol{r}_j |^3} - \sum_{j=1}^N 
     {G M_j \boldsymbol{r}_{j} 
     \over |\boldsymbol{r}_{j}|^3} +
     \boldsymbol{\Gamma}_{i}  \; ,
\label{emot}
\end{equation} 

\noindent where $G$ is the gravitational constant and $M_\star$, $M_i$
and $\boldsymbol{r}_i$ denote the mass of the central star, that of
core~$i$ and the position vector of core $i$, respectively.  The
third term on the right--hand side is the acceleration of the
coordinate system based on the central star (indirect term).

\noindent Acceleration due to tidal interaction with the disc is dealt
with through the addition of extra forces as in Papaloizou \& Larwood
(2000, see also Terquem \& Papaloizou 2007):

\begin{equation}
\boldsymbol{\Gamma}_{i} = 
-\frac{1}{t_{\rm m,i}} \frac{d \boldsymbol{r}_i}{dt} -
\frac{2}{| \boldsymbol{r}_i|^2 t_{\rm e,i}} 
\left( \frac{d  \boldsymbol{r}_i}{dt} \cdot
\boldsymbol{r}_i \right) 
\boldsymbol{r}_i - 
\frac{2}{ t_{\rm i,i}}
\left( \frac{d \boldsymbol{r}_i}{dt} \cdot {\bf e}_z \right) {\bf e}_z,
\end{equation}

\noindent where ${\bf e}_z$ is the unit vector perpendicular to the
disc midplane and $t_{\rm m,i}$, $t_{\rm e,i}$ and $t_{\rm i,i}$ are
the timescales over which, respectively, the angular momentum, the
eccentricity and the inclination with respect to the disc midplane of
the orbit of core $i$ change due to tidal interaction with the disc.
Note that the timescale on which the semimajor axis decreases is
$t_{\rm m,i}/2$ (e.g., Teyssandier \& Terquem 2014).  As here we are
not interested in following the evolution of a core after it gets
close to the star, we wo not include contribution from the tides
raised by the star nor from relativistic effects. 

\subsection{Type--I migration and collisions}

\label{sec:migration}

The cores we consider here are small enough that they undergo type~I
migration.   Radiation--hydrodynamical simulations of disc/planet
  interactions have shown that cores with masses between about 4 and
  30~$\ME$ and eccentricities below $\sim 0.015$ undergo outward
  migration, due to the effect of the corotation torque (Paardekooper
  \& Mellema 2006, Kley, Bitsch \& Klahr 2009, Bitsch \& Kley 2010).
  Planets more massive than about 30~$\ME$ open up a gap, which
  reduces the corotation torque, so that the total torque is negative
  and migration is inward.  However, as mentionned above, recent MHD
  simulations indicate that planets with masses significantly smaller
  (by at least a factor 3) than 30~$\ME$ can open up gaps in turbulent
  regions of discs with net vertical magnetic flux (Zhu, Stone \&
  Rafikov 2013).  Therefore, the range of planet masses for which
  outward migration occurs may be much smaller than suggested by the
  hydrodynamical simulations.  In this context, we will assume in this
  paper that type~I migration is always inward.  Note that our results
  would not be significantly affected if cores with masses in a narrow
  range and eccentricities below $\sim 0.015$ were migrating outward.

In the regime of inward type--I migration, Papaloizou \& Larwood
(2000) have shown that $t_{\rm m,i}$ and $t_{\rm e,i}$ can be written
as:

\begin{equation}
  t_{\rm m,i} = 146.0  \; \left[ 1+ \left( \frac{ e_i }{1.3 H/r}\right)^5 
\right]
  \left[ 1- \left( \frac{ e_i }{1.1 H/r}\right)^4 \right]^{-1} 
  \; \left( \frac{H/r}{0.05}
  \right)^2 \;
  \frac{{\rm M}_\odot}{M_d} \;
  \frac{{\rm M}_\oplus}{M_i} \; \frac{a_i}{{\rm 1~au}} \; \; \; {\rm years} ,
\label{tm}
\end{equation}

\begin{equation}
  t_{\rm e,i}^d = 0.362 \; \left[ 1+ 0.25 \left( \frac{ e_i }{H/r}\right)^3
  \right] \; \left( \frac{H/r}{0.05}
  \right)^4 \;
  \frac{{\rm M}_\odot}{M_d} \; \frac{{\rm M}_\oplus}{M_i} \;
  \frac{a_i}{{\rm 1~au}} \; \; \; {\rm years} ,
\label{te}
\end{equation}

\noindent and $t_{i,i}=t_{e,i}$.  Here $e_i$ is the eccentricity of
core~$i$, $H/r$ is the disk aspect ratio and $M_d$ if the disk mass
contained within 5~au.  The equations above assume that the disk
surface mass density varies like $r^{-3/2}$.

Collisions between cores are dealt with in the following way: if the
distance between cores~$i$ and~$j$ becomes less than $R_i+R_j$, where
$R_i$ and $R_j$ are the radii of the cores, a collision occurs and the
cores are assumed to merge.  They are subsequently replaced by a
single core of mass $M_i+M_j$ with the position and the velocity of
the center of mass of cores~$i$ and~$j$.

\subsection{Initial set up}

We start with a population of $N$ cores on circular orbits in the disc
midplane spread between an inner radius $R_{\rm in}$ and an outer
radius $R_{\rm out}$.  The initial distance between a core and the
star is chosen randomly.  The disc is assumed to be truncated at an
inner radius $R_{\rm cav}$, which in some simulations will increase
with time.

We assume that once a core reaches this radius $R_{\rm cav}$ it
  loses contact with the disc and stops migrating.  As indicated in
  section~\ref{sec:insitu}, it has been suggested that the cores may
  be trapped at the disc inner edge rather than penetrate inside the
  cavity.  When that happens, if the disc inner edge then expands, the
  planet may stay coupled to the disc and also move outward (Masset et
  al. 2006).  However, such a shepherding of the planet by the disc
  requires that the disc can tranfer enough angular momentum to the
  planet so that it can move outward as fast as the disc radius (Lyra
  et al. 2010).  This cannot be satisfied if X--ray photoevaporation
  is responsible for the expansion of the disc's inner cavity (Owen,
  Ercolano \& Clarke 2011), as the surface density of gas in the
  vicinity of the planet decreases to zero.  Therefore, in the
  simulations presented below, a planet reaching the disc inner radius
  will be assumed to decouple from the disc and will stay at its
  location when this radius moves out.

All the cores are supposed to have an identical mass density
$\rho=1$~g~cm$^{-3}$.  Note that this is smaller than the densities in
the Kepler--10 system, wich are inferred to be 5.8 and 7.1~g~cm$^{-3}$
for the 3 and 17~$\ME$ planets, respectively.  Therefore, the radii of
the cores in our simulations, which are given by $R_i=[3M_i/(4 \pi
\rho)]^{1/3}$, are almost twice as large as they would be if we
adopted those higher values of the density.  Thus, collisions between
cores are favoured in our model.  This, however, does not affect our
results, as we will find that collisions are not efficient enough for
the evolution of the population of cores to result in a 17~$\ME$ at
0.24~au.

In the simulations presented below, we have adopted $M_{\star}=1 \;
\Msun$, $M_d = 10^{-3} \; \Msun$ and $H/r=0.05$.  For these values of
the parameters, equations~(\ref{tm}) and~(\ref{te}) give $t_{\rm m,i}
\simeq 10^5 $~years and $t_{\rm e,i} \simeq 4 \times 10^2 $~years,
respectively, for a 1~$\ME$ planet on a circular orbit at 1~au.  

We now describe the results of our simulations. 
  
\subsection{A super Earth at 0.017~AU}

To investigate whether the dynamical evolution of a population of
migrating cores could result in a 3~$\ME$ planet at 0.017~au and a
17~$\ME$ planet at 0.24~au, and nothing else, we have run a series of
simulations with a total mass of cores equal to 20~$\ME$.  We have
considered cores with initial masses between 1 and 3~$\ME$, and $N$ in
the range 7 to 20.  In some simulations, all the cores have the same
mass, while in others, there is a mixture of different masses.  The
inner edge of the disc is taken to be $R_{\rm cav}=0.017$~au to start
with, and is moved up to 0.24~au after a total mass of cores of a few
$\ME$ has reached it.  The initial inner and outer radii of the
population of cores, $R_{\rm in}$ and $R_{\rm out}$, are in the range
0.1--3~au and 1--5~au, respectively.

\begin{figure}
\begin{center}
\includegraphics[scale=0.7]{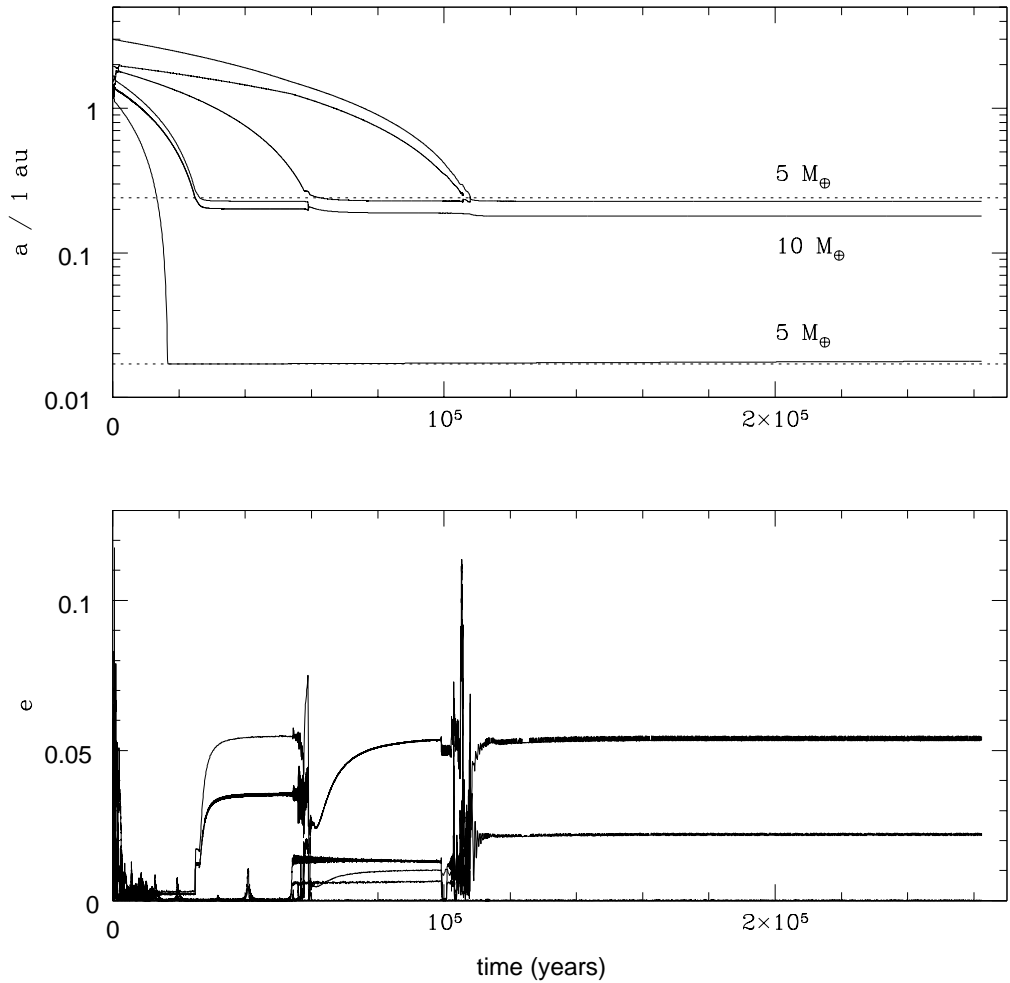}
\end{center}
\caption{Evolution of the semi--major axes (in units of au and in
  logarithmic scale; {\em upper plot}) and of the eccentricity ({\em
    lower plot}) of the 14 cores in the system versus time (in units
  of years).  Initially, the 6 outermost cores have a mass of 2~$\ME$,
  while the others have a mass of 1~$\ME$.  The solid lines correspond
  to the different cores.  A line terminates just prior to a
  collision.  On the upper plot, the dotted lines indicate the
  location of the inner cavity ($R_{\rm cav}=0.017$~au initially,
  0.24~au after $1.6 \times 10^4$~years). The disc is removed after
  $1.6 \times 10^5$~years.  There are 3 cores left at the end of the
  simulation.  Their masses are indicated on the upper plot.  }
\label{fig1}
\end{figure}

In figure~\ref{fig1}, we plot the results of a simulation with $N=14$
cores initially spread between $R_{\rm in}=1$~au and $R_{\rm
  out}=3$~au in a disc with an inner cavity below $R_{\rm
  cav}=0.017$~au.  The 6 outermost cores have a mass of 2~$\ME$, while
the 8 innermost cores have a mass of 1~$\ME$.  Very quickly after the
beginning of the simulation, a 5~$\ME$ core builds up through
collisions and migrates in.  It reaches the disc's inner cavity at
around $t=1.6 \times 10^4$~years, while the other cores are still
beyond 0.5~au.  After that time, the radius of the inner cavity is
moved up to 0.24~au.  The other cores continue to migrate in, and at
around $t=2.5 \times 10^4$~years, three cores with masses 4, 4 and
2~$\ME$ reach the new inner cavity's radius $R_{\rm cav}=0.24$~au.  As
two last cores reach this radius at around $t=1.1 \times 10^5$~years,
collisions occur, and finally two cores with masses 5~$\ME$ and
10~$\ME$ are left at 0.22 and 0.18~au, respectively, in a 7:5 mean
motion resonance.  After $t=1.6 \times 10^5$~years, the disc is
removed to make sure the system is stable.  The two outer planets,
being in a resonance, have rather large eccentricities, on the order
of a few hundreths, whereas the innermost planet has an eccentricity
below $10^{-3}$.

In the simulation described above, the outer edge of the cavity was
assumed to move up rather quickly, on a timescale of $\sim
10^4$~years.  However, this timescale could be made longer by
decreasing the mass of the disc, so that migration would be slower, or
by starting the cores further away from the central star.

We have run 37 simulations with a total mass of cores of 20~$\ME$, an
initial $R_{\rm cav}=0.017$~au and various $R_{\rm in}$ and $R_{\rm
  out}$.  In 6 of these simulations, the eccentricity damping
timescale given by equation~(\ref{te}) was increased by a factor of 2
or 5 to allow eccentricities to reach higher values, which would
promote collisions.  In 5 of the simulations, the initial masses of
the cores were 3 or 4~$\ME$, while in all the others they were 1 or
2~$\ME$.

We have obtained a single core close to $R_{\rm cav}=0.017$~au in 7 of
these simulations.  The mass of this planet was 1, 5, 5, 10, 6, 4 or
8~$\ME$, with the three last cases corresponding to simulations with
increased eccentricity damping timescale.  An inner core with 1~$\ME$
was obtained when one core in the initial distribution was detached
from the rest of the population and closer in than the others.  In all
of the 6 other cases, the core that came to a halt close to 0.017~au
built up through collisions very early on in the simulations.  Being
heavier than the others, it then migrated in faster and reached the
inner edge of the disc before the other cores had time to join.

In the other 30 simulations, several cores of a few Earth masses ended
up in mean motion resonances close to 0.017~au.  In most cases,
the cores would grow on their way in, at the same time as they were
migrating.

These simulations therefore indicate that, if a single core of a
few~$\ME$ at 0.017~au has grown by collisions and mergers of smaller
cores, most likely it has assembled further away.  It grew and
detached itself from a population of other smaller cores at a
distance of at least a few~au from the central star.

\subsection{A massive planet at 0.24~AU}

We now investigate how a massive core  which comes to a halt at
0.24~au could have formed.
 
In the 7 simulations described above where a single core ended up
close to 0.017~au, the other cores would still be beyond 0.5~au when
the inner core reached $R_{\rm cav}$.  We therefore subsequently moved
$R_{\rm cav}$ up to 0.24~au to investigate whether a single other core
could be obtained at this location.  In none of these simulations did
we obtain a single other core.  At least two cores in mean motion
resonances were left close to 0.24~au, as observed in
figure~\ref{fig1}.

To study more generally whether a single core could grow through
collisions and mergers within a population of cores with a total mass of
17~$\ME$, we performed another 29 simulations starting with cores with
masses between 1 and 3~$\ME$, $N$ in the range 6 to 17 and $R_{\rm
  cav}=0.24$~au initially.  The initial inner and outer radii of the
population of cores, $R_{\rm in}$ and $R_{\rm out}$, were in the range
1--3~au and 2--5~au, respectively.  In 6 of the simulations, the
initial spacing between two cores was set to be 4 or 4.5 times their
mutual Hill radius (as in Pierens et al. 2013).  In all the other
simulations, the location of the cores was chosen randomly between
$R_{\rm in}$ and $R_{\rm out}$.  Migration and eccentricity damping
timescales were computed from equations~(\ref{tm}) and~(\ref{te}).  In
10 cases, the simulation ended with two cores in mean motion resonance
close to the disc inner edge.  In the other cases, there were at least
3 cores left.  None of the simlulations ended with only one core.

\begin{figure}
\begin{center}
\includegraphics[scale=0.7]{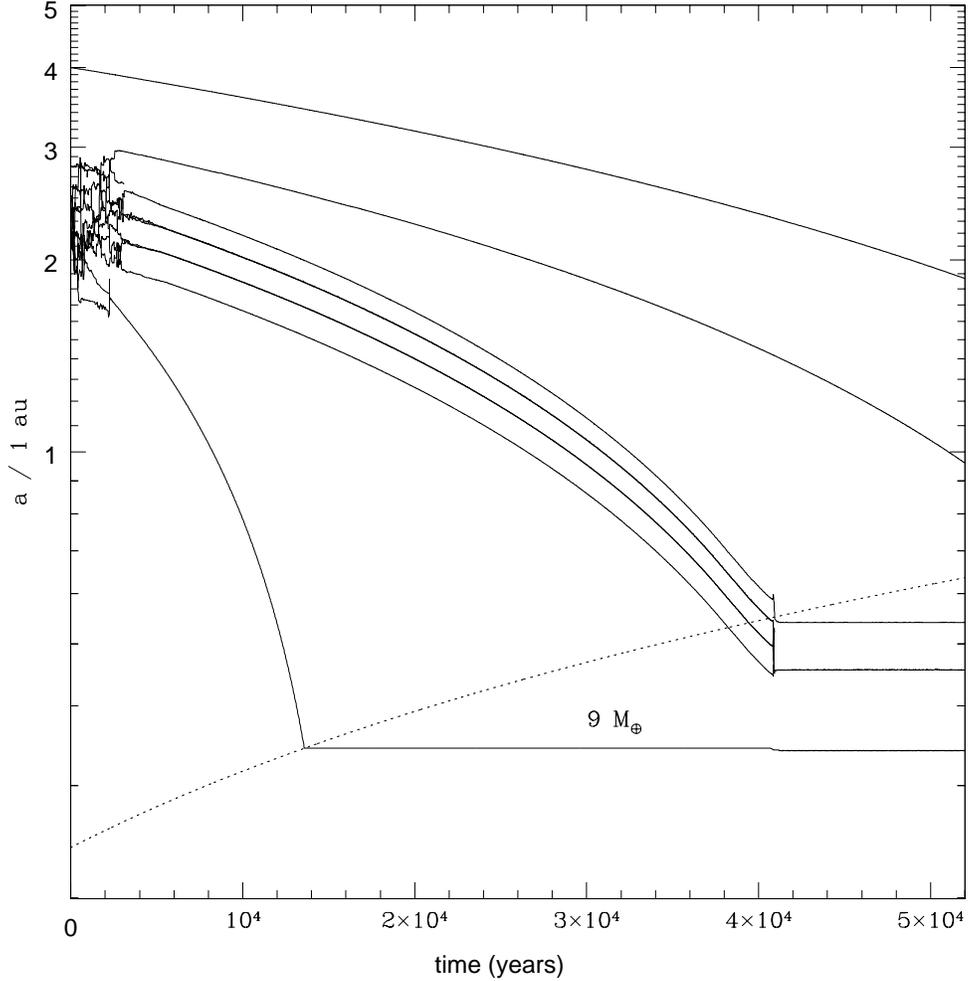}
\end{center}
\caption{Evolution of the semi--major axes (in units of au and in
  logarithmic scale) of the 14 cores in the system versus time (in
  units of years). Initially, the 5 outermost cores have a mass of
  3~$\ME$, the innermost core has a mass of 1~$\ME$ and the others
  have a mass of 2~$\ME$.  The solid lines correspond to the different
  cores.  A line terminates just prior to a collision.  The dotted
  line indicates the location of the inner cavity.  At the end of the
  simulation, there is a 9~$\ME$ core at 0.34~au, 2 cores in mean
  motion resonance close to 0.5~au and still two cores between 1 and
  2~au migrating in.}
\label{fig2}
\end{figure}

We then performed another 14 simulations with a larger total mass of
cores, to study whether a massive core could build--up through
collisions and migrate quickly to the inner edge before the others had
time to join.  In some of the simulations, the edge of the outer
cavity was assumed to increase linearly with time so that $R_{\rm
  cav}=1$~au after $10^5$~years.  In 2 of the simulations, we obtained
a rather massive core (9 or 10~$\ME$) at around 0.3~au.  In
figure~\ref{fig2}, we plot the results of one of these simulations.
It starts with $N=14$ cores spread between 2 and 4~au.  Initially, the
5 outermost cores have a mass of 3~$\ME$, the innermost core has a
mass of 1~$\ME$ and the others have a mass of 2~$\ME$, so that the
total mass is 32~$\ME$. The edge of the inner cavity starts at $R_{\rm
  cav}=0.24$~au and increases to 1~au after $10^5$~years.  We
terminate the simulation after $5 \times 10^4$~years, when there is a
9~$\ME$ core at 0.34~au, 2 cores in mean motion resonance close to
0.5~au and still two cores between 1 and 2~au migrating in.

The simulation described above results in a core less massive than the
one detected in the Kepler~10 system at 0.24~au, and there are two
other massive cores rather near by.  However, it does illustrate that
it is possible to get a massive core at a few tenths of an au starting
with a massive population of cores further away.  The mass of the core
reaching the inner edge could be increased by increasing the total
mass of the population of cores.  Also, if it grew further away from
the central star and detached itself from the rest of the population,
it would reach the inner edge while the other cores would still be far
away, so that at the end of the evolution no other core would be found
near by.  Note that the timescale over which the edge of the cavity is
moved is rather fast, so that we could perform a large number of
simulations, but again this timescale could be made longer by starting
the cores further away.

Here again, we note that the core that comes to a halt at around
0.3~au has assembled very early on in the simulation, at a distance of 
$\sim$~1~au from the central star.

\section{Critical core mass}
\label{sec:criticalmass}

The results presented in the previous section indicate that the
planets have formed at a distance of at least a few~au from the
central star before migrating in.  We therefore calculate what the
critical core mass is at this location and all the way down to
0.24~au.  Because the planets in the Kepler--10 system are very dense,
they have not accreted much gas, and therefore should not have
attained the critical core mass (see the discussion at the end of
section~\ref{sec:calculations}).  In the section below, we study the
conditions which are required for the critical core mass to be above
17~$\ME$ within a distance of a few au from the central star.

\subsection{Structure of the protoplanet atmosphere}
\label{sec:envelope}

Because the critical core mass corresponds to the mass of the core
above which no atmosphere can exist at equilibrium around it, we
solve the equations describing an atmosphere at equilibrium as a
function of the core mass.  The critical core mass is reached when these
equations no longer have a solution.

The equations governing the structure of the protoplanet atmosphere at
hydrostatic and thermal equilibrium  have been
presented in Papaloizou \& Terquem (1999) and we recall them below.

We assume that the protoplanet is spherically symmetric and
nonrotating.  We denote $\varpi$ the radius in spherical coordinates in a
frame with origin at the centre of the protoplanet.  The equation of
hydrostatic equilibrium is:

\begin{equation}
\frac{dP}{d \varpi } = - g \rho .
\label{dpdvarpi}
\end{equation}

\noindent Here, $P$ is the pressure, $g=G M(\varpi) / \varpi^2$ is the
acceleration due to gravity, with $M(\varpi)$ being the mass interior to
radius $\varpi$ (this includes the core mass if $\varpi$ is larger than the core
radius) and $G$ is the gravitational constant.  The mass
$M(\varpi)$ is related to the mass density per unit volume $\rho$ through:

\begin{equation}
\frac{dM}{d \varpi} = 4 \pi \varpi^2 \rho.
\label{dmdvarpi}
\end{equation}

We use the equation of state for a hydrogen and helium mixture given
by Chabrier et al. (1992) for mass fractions of hydrogen and helium of
0.7 and 0.28, respectively.  The luminosity $L_{\rm rad}$ that is
transported by radiation through the atmosphere is related to the
temperature gradient $dT/d \varpi$ through the standard equation of
radiative transport:

\begin{equation}
\frac{dT}{d \varpi} = \frac{-3 \kappa \rho}{16 \sigma
T^3} \frac{L_{\rm rad}}{4 \pi \varpi^2} ,
\label{dtdvarpi}
\end{equation}

\noindent where $\kappa$ is the opacity, which in
general depends on both $\rho$ and $T$, and $\sigma$ is the
Stefan--Boltzmann constant.

The total luminosity is transported by both radiation (in the outer
parts of the atmopshere) and convection (in the inner parts).  Here,
the only energy source for the atmopshere that we consider comes from
the planetesimals that are accreted by the protoplanet and release
their gravitational energy as they collide with the surface of the
core.  The corresponding total core luminosity $L_c$ is:

\begin{equation}
L_c = \frac{ G M_{c} \dot{M}_{c}}{ r_{c}} ,
\end{equation}

\noindent where $M_{c}$ and $r_{c}$ are, respectively, the mass and
the radius of the core, and $\dot{M}_{c}$ is the planetesimal
accretion rate.  

The radiative and adiabatic temperature gradients,
$\nabla_{\rm rad}$ and $\nabla_{\rm ad}$, are given by:

\begin{equation}
\nabla_{\rm rad} = \left( \frac{\partial \ln T}{\partial \ln P}
\right)_{\rm rad} = \frac{3 \kappa L_c P}{64 \pi
\sigma G M T^4} ,
\label{dTdr_rad}
\end{equation}

\noindent and

\begin{equation}
\nabla_{\rm ad} = \left( \frac{\partial \ln T}{\partial \ln P} \right)_{s} ,
\end{equation}

\noindent where the subscript ${s}$ indicates that the derivative has
to be evaluated at constant entropy.

\noindent When $\nabla_{\rm rad} < \nabla_{\rm ad}$,
there is stability to convection and therefore all the energy is
transported by radiation, i.e. $L_{\rm rad}=L_c$.  In the regions
where $\nabla_{\rm rad} > \nabla_{\rm ad}$, there is instability to
convection and therefore part of the energy is transported by
convection, i.e.  $L_c= L_{\rm rad}+L_{\rm conv}$, where
$L_{\rm conv}$ is the luminosity associated with convection.  
Using the mixing length theory (Cox~\& Giuli 1968), we obtain:


\begin{equation}
  L_{\rm conv}  =  \pi \varpi^2 C_p \Lambda_{\rm ml}^2 
  \left[ \left( \frac{\partial
        T}{\partial \varpi} \right)_{s} - 
    \left( \frac{\partial T}{\partial \varpi}
    \right) \right]^{3/2}
  \sqrt{ \frac{1}{2} \rho g \left| \left(
        \frac{\partial \rho}{\partial T} \right)_P \right| } ,
\end{equation}

\noindent where $\Lambda_{\rm ml}=|\alpha_{ml}P/(dP/d \varpi)|$ is the
mixing length, $\alpha_{ml}$ being a constant of order unity, $\left(
\partial T/\partial \varpi \right)_{s} = \nabla_{\rm ad} T \left( d \ln
P / d \varpi \right)$, and the subscript $P$ denotes evaluation at constant
pressure.  The different thermodynamic parameters needed in the above
equation are given by Chabrier et al. (1992), and we fix
$\alpha_{\rm ml}=1$.

\subsection{Boundary conditions}
\label{sec:boundary}

As we solve the above equations for the three variables $P$, $M$ and
$T$ as a function of $\varpi$, we need three boundary conditions.

We take for the mass density of the core $\rho_{c} = 7$~g~cm$^{-3}$,
which is approximately the value inferred for the 17~$\ME$ planet in
the Kepler--10 system (Dumusque et al. 2014).

We can then calculate the inner boundary of the atmosphere, which is
equal to the core radius $r_c$, given by:
\begin{equation}
r_{c} = \left( \frac{3 M_{c}}{4 \pi \rho_{c}} \right)^{1/3}.
\label{eq:rc}
\end{equation}
The first boundary condition is that $M(r_c)=M_{c}.$

The outer boundary of the atmosphere is taken to be at the Roche lobe
radius $r_L$ of the protoplanet, which is given by:

\begin{equation}
r_L = \frac{2}{3} \left( \frac{\Mp}{3 \Ms} \right)^{1/3} r ,
\end{equation}

\noindent where $\Mp = M_c + M_{\rm atm}$ is the planet mass,
$M_{\rm atm}$ being the mass of the atmosphere, and $r$ is the orbital
radius of the protoplanet in the disc.

We denote the disc midplane temperature,
pressure and mass density at the distance $r$ from the central star by
$T_m,$ $P_m$ and $\rho_m$, respectively.

\noindent At $\varpi=r_L$, we have $M(r_L)=\Mp$ and the two boundary
conditions $P=P_m$ and  $T$ given by:

\begin{equation}
T = \left( T_m^4 + \frac{ 3\tau_L L_c}{16 \pi \sigma r_L^2}
\right)^{1/4}.
\label{eq:Tbound}
\end{equation}

\noindent  This equation expresses the fact that the radiative
  flux at the surface of the protoplanet, $\sigma T^4$, is the sum of
  the radiative flux coming from the disc above the protoplanet,
  $\sigma T_m^4$, and the radiative flux coming from inside the
  protoplanet, $3\tau_L L_c / (16 \pi r_L^2)$.  This latter term takes
  into account the fact that the luminosity escaping from the surface
  of the protoplanet, $L_c$, is radiated after passing through an
  additional optical depth $\tau_L$ above the protoplanet atmosphere.
  In other words, $T$ must be larger than $T_m$ at $\varpi=r_L$ for
  the luminosity to be radiated away from the protoplanet into the
  surrounding disc.  We approximate $\tau_L$ by:

\begin{equation}
\tau_L = \kappa \left( \rho_m, T_m \right) \rho_m r_L .
\end{equation}

\noindent  As pointed out by Papaloizou \& Terquem (1999), the
  structure of the atmosphere is sensitive to the value of $T$ at
  $\varpi=r_L$ only when a significant part of the envelope is
  convective.  This occurs in the hot inner parts of the disc, below
  $\sim 0.1$~au.  Therefore, at the location of the 17~$\ME$ in the
  Kepler--10 system and beyond, the critical core mass is not
  sensitive to the boundary condition given by
  equation~(\ref{eq:Tbound}).

\subsection{Kelvin--Helmholtz timescale}

For a fixed $\dot{M}_c$ and at a given radius $r$, there is a critical
core mass $M_{\rm crit}$ above which no solution to the above
equations can be found.  As long as $M_c<M_{\rm crit}$, the energy
lost by the envelope through radiation is compensated for by the
gravitational energy which the planetesimals entering the atmosphere
release when they collide with the surface of the core.  The
atmosphere is then in quasi--static and thermal equilibrium.  However,
when $M_c > M_{\rm crit}$, the atmosphere can no longer be supported
at equilibrium.  It has to contract gravitationally to supply part of
the energy which is radiated away.  Rapid accretion of the gas in the
surrounding nebula then occurs.

How fast this accretion process is depends on how fast the envelope
can radiate away the energy which is produced by its gravitational
collapse.  This is given by the Kelvin--Helmholtz timescale, which can
be estimated as:

\begin{equation}
t_{\rm KH}= \frac{|E_T|}{L_c},
\label{eq:tKH}
\end{equation}

\noindent where $E_T$ is the total internal and gravitational energy
of the gas in the atmosphere when the core reaches the critical mass.
The luminosity that appears in equation~(\ref{eq:tKH}) is $L_c$ as
this is roughly the luminosity of the core when it becomes critical.

\subsection{Calculations}

\label{sec:calculations}

We compute the disc midplane temperature $T_m$, pressure $P_m$ and
mass density $\rho_m$ assuming a standard steady--state $\alpha$ disc
model (see Papaloizou \& Terquem 1999 for the details of the
computation).  Such a model is completely characterized by two
parameters, which we take to be $\alpha$ and the gas accretion rate
$\dot{M}_{\rm gas}$ through the disc.

For a particular disc model, at a fixed radius $r$ in the disc, for a
given core mass $M_c$ and planetesimal accretion rate $\dot{M}_c, $ we
solve equations~(\ref{dpdvarpi}), (\ref{dmdvarpi})
and~(\ref{dtdvarpi}) with the boundary conditions described above to
get the structure of the envelope.  The opacity is taken from Bell~\&
Lin (1994) and has contributions from dust grains, molecules, atoms
and ions.  
The value of $M_c$ above which the equations have no solution is the
critical core mass $M_{\rm crit}$.

In table~\ref{table1}, we give the values of $M_{\rm crit}$,
$\Mp=M_{\rm crit} + M_{\rm atm}$ and of the Kelvin--Helmholtz
timescale $t_{\rm KH}$ for disc models with $\dot{M}_{\rm
  gas}=10^{-8}$~$\Msun$~yr$^{-1}$ and $\alpha=10^{-3}$ or $10^{-2}$,
at the radii $r=0.24$ and 1~au in the disc, and for a planetesimal
accretion rate $\dot{M}_c=10^{-7}$, $10^{-6}$ or $10^{-5}$~$\mey$.  By
comparing $\Mp$ and $M_{\rm crit}$ we see that, when the core reaches
the critical mass, $\Mp \simeq 1.5 M_{\rm crit}$ (in agreement with
Bodenheimer \& Pollack 1986).

\begin{table}
\caption{Critical core mass and Kelvin--Helmholtz timescale}
\begin{tabular}{cccccccc}
\hline
$\alpha$ & $r$ & $T_m$ & $P_m$ & $\dot{M}_c$ & $M_{\rm crit}$ &
$\Mp$ & $t_{\rm KH}$ \\
  & (au)& (K)   &  (erg~cm$^{-3}$) & ($\mey$) & ($\ME$) &
($\ME$) & ($10^6$ yr) \\
\hline
$10^{-2}$ & 0.24 & 1001.1 & 41.0 & $10^{-5}$ & 25.8 & 38.9 & 0.16 \\
 -- & -- & -- & -- & $10^{-6}$ & 18.4 & 27.4 & 1.3 \\
 -- & -- & -- & -- & $10^{-7}$ & 13.1 & 19.5 & 11.0 \\
-- & 1 & 273.0 & 1.1 &  $10^{-5}$ & 24.3 & 36.3 & 0.13 \\
-- & -- & -- & -- &  $10^{-6}$ & 16.7 & 24.6 & 0.96 \\
-- & -- & -- & -- &  $10^{-7}$ & 11.2 & 16.6 & 7.1 \\
\hline
$10^{-3}$ & 0.24 & 1180.8 & 359.8 & $10^{-5}$ & 24.6 & 36.2 & 0.14 \\
 -- & -- & -- & -- & $10^{-6}$ & 17.5 & 26.4 & 1.2 \\
 -- & -- & -- & -- & $10^{-7}$ & 12.5 & 18.7 & 9.9 \\
-- & 1 & 480.7 & 8.5 &  $10^{-5}$ & 23.9 & 35.6 & 0.12 \\
-- & -- & -- & -- &  $10^{-6}$ & 16.6 & 24.6 & 0.95 \\
-- & -- & -- & -- &  $10^{-7}$ & 11.4 & 16.5 & 7.3 \\
\hline
\end{tabular} \\
\medskip \\
Listed are the parameter $\alpha$ used in the disc models (column~1), 
the orbital 
radius $r$ of the core in~au (column~2), the disc midplane temperature in~K 
(column~3) and pressure in erg~cm$^{-3}$ (column~4) at this radius, the 
planetesimal 
accretion rate onto the core  $\dot{M}_c$ in $\mey$ (column~5), the 
critical core 
mass $M_{\rm crit}$ in~$\ME$ (column~6), 
the total mass of the planet $\Mp=M_{\rm crit} + M_{\rm atm}$ in~$\ME$ 
(column~7) 
and the  Kelvin--Helmholtz timescale 
$t_{\rm KH}$ in Myr for a core with the critical mass (column~8).
\label{table1}
\end{table}

\begin{figure}
\begin{center}
\includegraphics[scale=0.7]{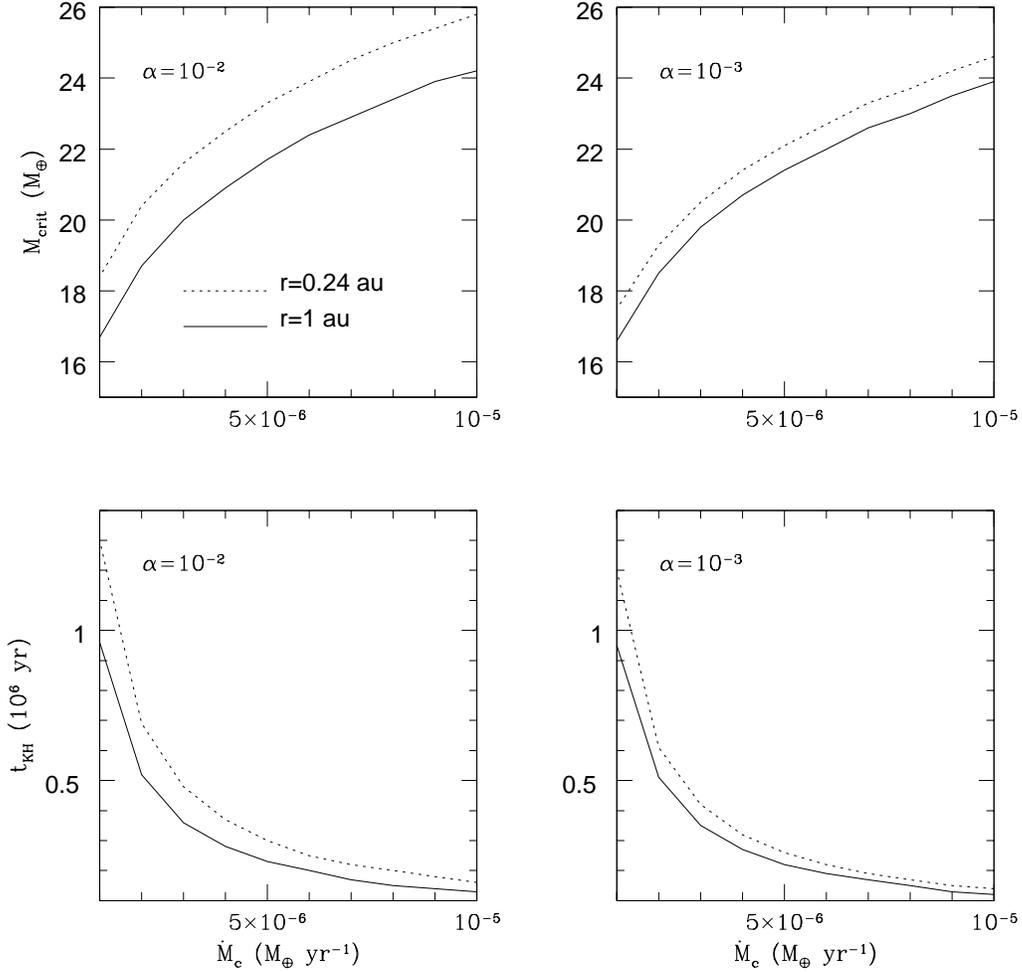}
\end{center}
\caption{Critical core mass $M_{\rm crit}$ in units of~$\ME$ (upper
  panels) and Kelvin--Helmholtz timescale $t_{\rm KH}$ in units of
  $10^6$~yr for a core with the critical mass (lower panels) as a
  function of the planetesimal accretion rate onto the core
  $\dot{M}_c$ in $\mey$ at $r=1$~au (solid lines) and $r=0.24$~au
  (dotted lines) for a disc model with $\alpha=10^{-2}$ (left panels)
  and $10^{-3}$ (right panels). The values of $M_{\rm crit}$ beyond
  1~au are roughly the same as at 1~au. }
\label{fig3}
\end{figure}

Figure~\ref{fig3} shows $M_{\rm crit}$ and $t_{\rm KH}$ as a function
of $\dot{M}_c$ in the range $10^{-6}$--$10^{-5}$~$\mey$ at $r=0.24$
and 1~au and for disc models with $\dot{M}_{\rm
  gas}=10^{-8}$~$\Msun$~yr$^{-1}$ and $\alpha=10^{-3}$ or $10^{-2}$.

As was already noted by Papaloizou \& Terquem (1999), $M_{\rm crit}$
is essentially independant of $r$ for $r$ larger than about 0.1~au.
This is because $M_{\rm crit}$ depends on the boundary conditions only
when a large part of the envelope is convectively unstable, which
happens only for the highest values of $T_m$ and $P_m$, i.e. in the
disc's inner parts.  The values of $M_{\rm crit}$ beyond 1~au can
therefore be taken as being roughly the same as at 1~au.

From table~\ref{table1} and figure~\ref{fig3}, we see that $\dot{M}_c$
has to be larger than $10^{-6}$~$\mey$ for $M_{\rm crit}$ to be larger
than 17~$\ME$ beyond 0.24~au.  For such values of $\dot{M}_c$, a
$17$~$\ME$ core forming at a few~au from the star and migrating in
would not be expected to accrete a massive atmosphere of gas.
However, the core could still accrete an envelope that would stay at
equilibrium at its surface.  The mass of an envelope at equilibrium
onto a 17~$\ME$ core depends on $\dot{M}_c$.  The largest value is
attained when the core is very close to being critical, and in that
situation $\Mp= M_{\rm crit}+M_{\rm atm} \simeq 1.5 M_{\rm crit}$,
which gives $M_{\rm atm}=8.5$~$\ME$.  From table~\ref{table1}, we see
that a 17~$\ME$ core is close to being critical if
$\dot{M}_c=10^{-6}$~$\mey$, and the corresponding Kelvin--Helmholtz
timescale is $t_{\rm KH} \simeq 10^6$~years at $r \ge 0.24$~au.  If
$\dot{M}_c=10^{-5}$~$\mey$, we calculate that the mass of the
atmosphere at equilibrium onto a 17~$\ME$ core is much smaller, being
$M_{\rm atm} \simeq 1$~$\ME$, and for such an atmosphere $t_{KH}
\simeq 2 \times 10^4$~yr at $r \ge 0.24$~au in a disc with either
$\alpha=10^{-2}$ or $\alpha=10^{-3}$.

Therefore, if $\dot{M}_c=10^{-6}$~$\mey$, as the Kelvin--Helmholtz
timescale is much longer than the migration timescale, the core may
not have had time to accrete the 8.5~$\ME$ of gas that could be
supported at equilibrium before it reached the disc's inner cavity.
In contrast, if $\dot{M}_c=10^{-5}$~$\mey$, the Kelvin--Helmholtz
timescale is much shorter than the migration timescale, so the core
can accrete the whole atmosphere that can supported at equilibrium,
but that would only be about 1~$\ME$.  Therefore, in both cases, we
may expect an atmosphere at most on the order of an Earth mass on top
of the core.

As this atmosphere is not detected today, it has been stripped away.
Let us first show that Jean's escape at 0.24~au from the central star
cannot account for the disappearance of the atmopshere.  The escape
velocity from a core with mass $M_c$ and radius $r_c$ is $v_{\rm
  esc}=\left( 2 G M_c / r_c \right)^{1/2}$.  With $M_c=17$~$\ME$ and
$r_c$ given by equation~(\ref{eq:rc}), in which we take
$\rho_c=7$~g~cm$^{-3}$, we obtain $v_{\rm esc} \simeq 3 \times
10^4$~m~s$^{-1}$.  As the luminosity of the star in the Kepler--10
system is similar to that of the Sun, the temperature of the planet
atmosphere due to stellar irradiation, after the disc has disappeared,
is $T=\left[ L_\odot / \left( 4 \pi \sigma r^2 \right) \right]^{1/4}$,
where $r$ is the distance between the star and the planet.  As this
assumes that the atmosphere behaves like a blackbody, the derived
temperature is only a crude estimate.  At $r=0.24$~au, we obtain $T
\simeq 574$~K. This gives the thermal velocity of a hydrogen molecule,
$v_{\rm th} = \left( kT / m_p \right)^{1/2} \simeq 2 \times
10^3$~m~s$^{-1}$, where $k$ is the Boltzmann constant and $m_p$ is the
mass of the proton.  As $v_{\rm th}$ is an order of magnitude smaller
than $v_{\rm esc}$, Jean's escape cannot have operated for the
17~$\ME$ core at 0.24~au.  An alternative for stripping away the
atmosphere would be stellar wind (as as been proposed for Mars), giant
impacts or planetesimal accretion (see Schlichting, Sari \& Yalinewich
2014 and references therein) or mass loss due to the stellar XUV flux 
(Rogers et al. 2011).

In the above discussion, we have assumed that the mass of the
  planet had to be smaller than the critical core mass for a large
  quantity of gas not to be accreted.  In principle though, the planet
  could be more massive than the critical mass if the
  Kelvin--Helmholtz timescale were longer than the migration
  timescale.  The planet would then reach the disc inner edge and lose
  contact with the disc before a significant amount of gas could be
  accreted.  We now brifely show that this actually cannot be
  achieved.  If the planetesimal accretion rate were
  $\dot{M}_c=10^{-7}$~$\mey$, the critical core mass at 1~au would be
  about 11~$\ME$. The Kelvin--Helmholtz timescale onto a core reaching
  that mass being $\sim 7 \times 10^6$~years, such a core would enter
  the disc inner cavity without having accreted a significant amount
  of gas.  However, in the case of the Kepler--10 system, the core
  would have to grow up to 17~$\ME$ before reaching the disc inner
  edge.  A core of that mass embedded in a disc with
  $\dot{M}_c=10^{-7}$~$\mey$ has an atmosphere which cannot be at
  equilibrium, and which therefore is detached from the Roche lobe.
  Papaloizou \& Nelson (2005) have computed the evolution of a core
  embedded in a disc and which atmosphere is detached from the Roche
  lobe.  They found that such a protoplanet can accrete gas at any
  rate that may be supplied by the disc without expansion.  Therefore,
  for typical gas accretion rates, a significant atmosphere would be
  accreted onto the core before it entered the disc inner cavity.

\section{Summary and discussion}
\label{sec:discussion}

The simulations we have performed indicate that the planets in a
system like Kepler--10 have formed much further away from the central
star than the location at which they are detected today.  They cannot
have assembled through collisions and mergers of a population of low
mass cores with a total mass of 20~$\ME$ migrating in.  This is
because the eccentricity damping timescale is much shorter than the
migration timescale, so that the cores in such a population end up in
a resonant chain rather than collide which each other until only two
cores are left at 0.017 and 0.24~au.

Either (i) the planets grew all the way up by accreting planetesimals,
or (ii) they grew through collisions among a population of cores.  In
the first case, they had to gain their mass on a timescale shorter
than the migration timescale.  In the second case, they had to grow
fast enough that they would detach themselves from the population of
remaining cores (which total mass had to be significantly larger than
the mass of the two planets) and migrate in to the disc's inner edge
faster than the other, less massive cores.  By the time the other
cores migrate in significantly, the inner edge of the disc has moved
out, so that these cores are further away and cannot be detected.  In
this situation, the 3~$\ME$ core would have formed earlier on and/or
closer to the central star than the 17~$\ME$, so that the inner edge
of the disc would have had time to move from 0.017~au to 0.24~au out
in between their respective arrival in the disc's cavity.

In both cases, the planets have essentially acquired their mass at a
distance of at least a few~au from the central star.  The physical
conditions at this location are then relevant to study the accretion
of an atmosphere onto the cores.

As pointed out in section~\ref{sec:migration}, we have assumed
  that the cores, starting from the initial population, always migrated
  inward.  More specifically, to form a planetary system like
  Kepler--10, we need the 3 and 17~$\ME$ cores to migrate inward
  starting at a distance of at least a few~au.  According to the
  radiation--hydrodynamical simulations of disc/planet interactions
  (Bitsch \& Kley 2010), the 17~$\ME$ core would be expected to
  migrate outward, as its eccentricity is damped below $\sim 0.015$ by
  the interaction with the disc.  Our results therefore give support
  to the MHD simulations (Zhu et al. 2013) which show that a 17~$\ME$
  may open up a gap in a turbulent disc with a net vertical magnetic
  flux, thus reducing the contribution of the corotation torque and
  enabling inward migration. 

As the 17~$\ME$ planet in the Kepler--10 system is very dense and
probably does not have an atmosphere (Dumusque et al. 2014), it has
not reached the critical mass.  We have found that this requires the
planetesimal accretion rate onto the core to be larger than
$10^{-6}$~$\mey$.  This value, although in the upper range, is not
unphysical and has commonly been used in studies of planet formation
(Tanaka \& Ida 1999, Ikoma, Nakazawa \& Emori 2000 and references
therein).  A rather high value of the planetesimal accretion rate
during the planet formation phase is also consistent with the
existence of two rather massive solid planets in the Kepler--10
system, and suggests that this system has formed in a somewhat massive
disc.  If a core builds--up at a few au from the central star and
migrates in on a timescale of $\sim$~$10^5$~years, it would accrete
only about 0.1~$\ME$ of solid material on its way in if the
planetesimal accretion rate is uniform and equal to $10^{-6}$~$\mey$.
As the critical core mass does not depend much on the distance from
the central star beyond $\sim$~0.1~au, the core would therefore remain
subcritical.  If the planetesimal accretion rate were $10^{-5}$~$\mey$
instead, the core would have built--up to about 16~$\ME$ at a few au
from the central star and grown to its present mass on its way in.  In
that case, its mass would be much smaller than the critical mass.

Even a subcritical core can accrete a gaseous envelope, which stays at
quasi equilibrium around it.  We have found that, for a planetesimal
accretion rate between $10^{-6}$ and $10^{-5}$~$\mey$, the core would
have accreted an envelope of at most $\sim$~1~$\ME$.  This envelope
must have been stripped away as it is probably not present today.

The results presented in this paper indicate that a planetary system
like Kepler--10 may not be unusual, although it has probably formed in
a rather massive disc.  It is interesting to note that the
observations of both gas giant planets and massive solid planets are
consistent with the initial disc mass being a key parameter in
determining the final outcome of planetary systems.  Massive discs
favour the formation of massive planets which migrate in fast and end
up on short orbits  (as seen in the simulations by Thommes,
  Matsumura \& Rasio 2008).  However, gas giant planets may not
necessarily form in those discs if the planetesimal accretion rate is
high enough that even rather massive cores remain subcritical.

\section*{Acknowledgements}

I thank an anonymous referee for helpful comments and suggestions
that improved the manuscript.  It is a pleasure to thank J. Papaloizou for
stimulating discussions about gas accretion onto protoplanetary cores.

%
%


\begin{thebibliography}{}


\bibitem[2011]{B11} Batalha N. M., Borucki W. J., Bryson S. T.  {\em
    et al.}, 2011, ApJ, 729, 21

\bibitem[2013]{Batalha} Batalha N. M., Rowe J. F., Bryson S. T. 
{\em et al.}, 2013, ApJS, 204,24

\bibitem[1994]{Bell} Bell K. R., Lin D. N. C., 1994, ApJ, 427, 987

\bibitem[2010]{BK} Bitsch B., Kley W., 2010, A\&A, 523, 30

\bibitem[1986]{BP86} Bodenheimer P., Pollack J. B., 1986, Icarus, 67,
  391

\bibitem[2011]{Borucki} Borucki W. J., Koch D. G., Basri G. 
{\em et al.}, 2011, ApJ, 736, 19

\bibitem[2014]{Bucchave} Buchhave L. A., Bizzarro M., Latham D. W.
  {\em et al.}, 2014, arXiv:1405.7695

\bibitem[1992]{Chabrier} Chabrier G., Saumon D., Hubbard W. B., 
  Lunine J. I., 1992, ApJ, 391, 817

\bibitem[1968]{Cox}
Cox J. P., Giuli R. T., 1968, Principles of Stellar
Structure: Physical Principles (New York: Gordon \& Breach)


\bibitem[2014]{Dumusque} Dumusque X., Bonomo A. S., Haywood R. D. {\em
    et al.}, 2014, arXiv:1405.7881

\bibitem[2000]{Ikoma} Ikoma M., Nakazawa K., Emori H., 2000, ApJ, 537,
  1013

	
\bibitem[2009]{KBK} Kley W., Bitsch B., Klahr H., 2009, A\&A, 506, 971

\bibitem[2012]{LJ12} Lambrechts M., Johansen A., 2012, A\&A, 544, 32

\bibitem[2009]{Leger} L\'eger A., Rouan D., Schneider J. {\em et al.},
  2009, A\&A, 506,287

\bibitem[1993]{Lissauer} Lissauer J. J., 1993, ARA\&A, 31, 129

\bibitem[2010]{Lyra} Lyra W., Paardekooper S.--J., Mac Low M.--M.,
  2010, ApJ, 715, 68

\bibitem[2014]{Marcy} Marcy G. M., Weiss L. M., Petigura E. A. {\em et
    al.}, 2014, arXiv:1404.2960

\bibitem[2006]{MMCF} Masset F. S., Morbidelli A., Crida A., Ferreira
  J., 2006, ApJ, 642, 478

\bibitem[2011]{OEC} Owen J. E., Ercolano B., Clarke C. J., 2011,
  MNRAS, 412, 13

\bibitem[2006]{PM} Paardekooper S.--J., Mellema G., 2006, A\&A, 459,
  L17

\bibitem[2000]{PL00} Papaloizou J. C. B., Larwood J. D., 2000, MNRAS,
  315, 823

\bibitem[2005]{PN} Papaloizou J. C. B., Nelson R. P., 2005, A\&A,
  433, 247

\bibitem[1999]{PT99} Papaloizou J. C. B., Terquem C., 1999, ApJ, 521, 823

\bibitem[2001]{PT01} Papaloizou J. C. B., Terquem C., 2001, MNRAS,
  325, 221

\bibitem[2006]{PT06} Papaloizou J. C. B., Terquem C., 2006,
  Reports on Progress in Physics, 69, 119

\bibitem[2013]{Pierens} Pierens A., Cossou C., Raymond S. N., 2013,
  A\&A, 558, 14

\bibitem[2009]{Queloz} Queloz D., Bouchy F., Moutou C. {\em et al.},
  2009, A\&A, 506, 303

\bibitem[2011]{Rogers} Rogers L. A., Bodenheimer P., Lissauer J. J.,
  Seager S., 2011, ApJ, 738, 59

\bibitem[2014]{SSY} Schlichting H., Sari R., Yalinewich A., 2014,
  arXiv:1406.6435

\bibitem[1999]{Tanaka} Tanaka H., Ida S., 1999, Icarus, 139, 350

\bibitem[2008]{T08} Tanigawa T., 2008, P\&SS, 56, 1758

\bibitem[2007]{TP07} Terquem C., Papaloizou J. C. B.,  2007,
   ApJ, 654, 1110

\bibitem[2014]{TT14} Teyssandier J., Terquem C., 2014, arXiv:1406.2189

\bibitem[2008]{TMR} Thommes E. W., Matsumura S., Rasio F. A., 2008,
  Science, 321, 814

\bibitem[1997]{Ward} Ward W. R., 1997, ApJ, 482, 211

\bibitem[2013]{ZSR} Zhu Z., Stone J. M., Rafikov R. R., 2013, ApJ,
  768, 143

\end{thebibliography}

%

\label{lastpage}
\end{document}